\documentclass[journal]{IEEEtran}
\usepackage{textcomp}
\usepackage{graphicx}  
\graphicspath{{figures/}}
\usepackage{graphicx}
\usepackage{algorithm,algorithmic}
\usepackage{booktabs}
\usepackage{eurosym}
\usepackage{setspace}
\usepackage{amsmath,amssymb,amsthm,mathrsfs,amsfonts,dsfont} 
\usepackage[noadjust]{cite}
\makeatletter
\newcommand{\ALOOP}[1]{\ALC@it\algorithmicloop\ #1%
  \begin{ALC@loop}}
\newcommand{\ENDALOOP}{\end{ALC@loop}\ALC@it\algorithmicendloop}
\renewcommand{\algorithmicrequire}{\textbf{Input:}}

\makeatother

\begin{document}
\markboth{IEEE TRANSACTIONS ON SMART GRID}{}

\title{Residential Demand Response Applications\\ Using Batch Reinforcement Learning }
\author{F. Ruelens, B.J. Claessens,  S. Vandael, B. De Schutter, R. Babu\v{s}ka and R. Belmans.
\thanks{
F. Ruelens, S. Vandael and R. Belmans  are with the department of Electrical Engineering, KU~Leuven/EnergyVille, Leuven, Belgium. }
\thanks{
B.J. Claessens is with the energy department of VITO, Belgium.}
\thanks{
B. De Schutter  and R. Babu\v{s}ka are with Delft University of Technology, The Netherlands.}
\vspace{-4mm}}

\maketitle
\begin{abstract}

Driven by recent advances in batch Reinforcement Learning (RL), this paper contributes to the application of batch RL  to demand response. 
In contrast to conventional model-based approaches, batch RL techniques do not require a system identification step, which makes them more suitable for a large-scale implementation. 
This paper extends fitted Q-iteration, a standard batch RL  technique, to the situation where  a forecast of the exogenous data is provided.
In general, batch RL  techniques do not rely on expert knowledge on the system dynamics or the solution.
However, if some expert knowledge is provided, it can be incorporated by using our novel policy adjustment method.
Finally, we tackle the challenge of finding an open-loop schedule required to participate in the day-ahead market. 
We propose a model-free Monte-Carlo estimator method that uses a metric to construct artificial trajectories and we illustrate this method by finding the day-ahead schedule of a heat-pump thermostat. 
Our experiments show that batch RL  techniques  provide a valuable alternative to model-based controllers and that they can be used to  construct both closed-loop and open-loop policies.
\end{abstract}


\IEEEpeerreviewmaketitle

\newcommand{\ubar}[1]{\underaccent{\bar}{#1}}

\newcommand{\expected}{\mathbb{E}}
\newcommand{\uphk}{u_k^{\mathrm{ph}}}
\newcommand{\uph}{u^{\mathrm{ph}}}
\newcommand{\uphl}{u_l^{\mathrm{ph}}}
\newcommand{\xkt}{x_{k,\mathrm{t}}^{\mathrm{q}}}
\newcommand{\xkphys}{x_{k,\mathrm{ph}}}
\newcommand{\xkphysmin}{\underline{x}_{k,\mathrm{ph}}}
\newcommand{\xkphysmax}{\overline{x}_{k,\mathrm{ph}}}
\newcommand{\xphys}{x_{\mathrm{ph}}}
\newcommand{\xphysmin}{\underline{x}_{\mathrm{ph}}}
\newcommand{\xphysmax}{\overline{x}_{\mathrm{ph}}}
\newcommand{\umax}{u_{\mathrm{max}}}
\newcommand{\xlphys}{x_{l,\mathrm{ph}}}

\newcommand{\soc}{x_{\mathrm{soc}}}
\newcommand{\Tkin}{T_{k,\mathrm{in}}}
\newcommand{\Tkout}{T_{k,\mathrm{out}}}
\newcommand{\batchUphys}{ \mathcal{F} = \{(x_{l},u_{l},x_{l}^\prime,\uphl)\}_{l=1}^{\#\mathcal{F}}}
\newcommand{\batchCost}{ \mathcal{F} = \{(x_{l},u_{l},x_{l}^\prime,c_{l})\}_{l=1}^{\#\mathcal{F}}}

\newcommand{\tuple}{(x_{l},u_{l},x_{l}',\uphl)}

\newcommand{\forecastexo}{\hat{X}_{\mathrm{ex}}}
\newcommand{\forecastexoPhys}{\hat{X}_{\mathrm{ex}}^{\mathrm{ph}}}
\newcommand{\forecastexoCost}{\hat{X}_{\mathrm{ex}}^{\mathrm{c}}}

\newcommand{\probwk}{p_{\mathcal{W}}(\cdot|x_{k})}
\newcommand{\probw}{p_{\mathcal{W}}(\cdot|x)}
\newcommand{\Tin}{T_{\mathrm{i}}}
\newcommand{\Tint}{T_{\mathrm{i},k}}
\newcommand{\Tout}{T_{\mathrm{o}}}
\newcommand{\Toutt}{T_{\mathrm{o},k}}
\newcommand{\Hm}{H_{\mathrm{m}}}
\newcommand{\Cm}{C_{\mathrm{m}}}
\newcommand{\Ca}{C_{\mathrm{a}}}
\newcommand{\Ua}{U_{\mathrm{a}}}
\newcommand{\Tm}{T_{\mathrm{m}}}
\newcommand{\Qi}{Q_{\mathrm{i}}}
\newcommand{\Qg}{Q_{\mathrm{g}}}
\newcommand{\Qm}{Q_{\mathrm{m}}}
\newcommand{\Tindot}{\dot{T}_{\mathrm{i}}}
\newcommand{\Tmdot}{\dot{T}_{\mathrm{m}}}
\newcommand{\Qs}{Q_{\mathrm{s}}}
\newcommand{\Pheat}{P_{\mathrm{h}}}
\newcommand{\Pcool}{P_{\mathrm{c}}}
\newcommand{\Pheatmax}{P_{\mathrm{h}}^{\mathrm{max}}}
\newcommand{\Pcoolmax}{P_{\mathrm{c}}^{\mathrm{max}}}
\newcommand{\minimum}{\mathrm{max}}
\newcommand{\Uphys}{U^{\mathrm{ph}}}

\newcommand{\Qhp}{Q_{\mathrm{h}}}
\newcommand{\Paux}{P_{\mathrm{a}}}
\newcommand{\Tsett}{T_{\mathrm{s},k}}
\newcommand{\Tset}{\ubar{T}_{\mathrm{s}}}
\newcommand{\Tsetmax}{\bar{T_{\mathrm{s}}}}
\newcommand{\Taux}{T_{\mathrm{a}}}
\newcommand{\Tsetback}{T_{\mathrm{sb}}}
\newcommand{\Tbuffer}{T_{\mathrm{b}}}
\newcommand{\Tbufferaux}{T_{\mathrm{b}}^{\mathrm{a}}}

\newcommand{\nnim}{n_{\mathrm{min}}}
\newcommand{\Solart}{S_{k}}
\newcommand{\nsteps}{n_{\mathrm{s}}}
\newcommand{\Tintrace}{T_{\mathrm{i},k-1},...,T_{\mathrm{i},k-n}}
\newcommand{\utrace}{u_{k}^{\mathrm{ph}},u_{k-1}^{\mathrm{ph}},...,u_{k-n}^{\mathrm{ph}}}
\newcommand{\zenc}{z_{\mathrm{e}}}
\newcommand{\zenck}{z_{\mathrm{e},k}}

\newcommand{\Tsetvalue}{20.5^{\circ}\mathrm{C}}
\newcommand{\Tbuffervalue}{0.5^{\circ}\mathrm{C}}
\newcommand{\Tbufferauxvalue}{1.5^{\circ}\mathrm{C}}
\newcommand{\days}{8}
\newcommand{\atWork}{7}
\newcommand{\atHome}{17}
\newcommand{\result}{7\%}
\newcommand{\resultwinter}{8\text{-}9\%}
\newcommand{\resultsummer}{5\text{-}9\%}
\newcommand{\resultwinterkwh}{500\mathrm{kWh}}
\newcommand{\resultsummerkwh}{900\mathrm{kWh}}

\newcommand{\Xtime}{X_{\mathrm{t}}}
\newcommand{\Xphys}{X_{\mathrm{ph}}}
\newcommand{\Xexo}{X_{\mathrm{ex}}}

\newcommand{\Xtimequarter}{X_{\mathrm{t}}^{\mathrm{q}}}
\newcommand{\Xtimeday}{X_{\mathrm{t}}^{\mathrm{d}}}

\section{Introduction}
\IEEEPARstart{T}{he} increasing share of renewable energy sources introduces the need for flexibility on the demand side of the electricity system~\cite{Dupont}.
A prominent example of loads that offer flexibility at the residential level are thermostatically controlled loads, such as heat pumps, air conditioning units, and electric water heaters.
These loads represent about 20$\%$ of the total electricity consumption at the residential level in the United States~\cite{IEA}.
In addition, their market share is expected to increase as a result of the electrification of heating and cooling~\cite{IEA}, making them an interesting domain for optimization methods~\cite{koch2011modeling,Dupont,kara2012using,RuelensBRLCluster}.
Demand response programs offer  demand flexibility by motivating end users to adapt their consumption profile in response to changes in the electricity price or other grid signals.
The forecast uncertainty of  renewable energy sources~\cite{pinsonSG}, combined with their limited controllability,  have made demand response the topic of an extensive number of research projects~\cite{peeters2009address,Dupont,PowerMatchingCity} and scientific papers~\cite{koch2011modeling,TRIANA,oNeill2010residential,Vandael2012TSA,RuelensBRLCluster}. 
The traditional control paradigm defines the demand response problem as a model-based control problem~\cite{koch2011modeling,TRIANA,peeters2009address}, requiring a model of the demand response application, an optimizer, and a forecasting technique. 
A critical step in setting up a model-based controller comprises selecting accurate models and estimating the  model parameters.
This step becomes  more challenging considering the heterogeneity of the end users and their different patterns of behavior~\cite{challengesMPC}.
As a result, different end users are expected to have different model parameters and even different models.  
As such, a large-scale implementation of model-based controllers requires a stable and robust approach that is able to identify the appropriate model and the corresponding model parameters.

Reinforcement Learning (RL)~\cite{sutton1998reinforcement,bertsekas1996neuro}, on the other hand, is a model-free technique that requires no system identification step and no a priori knowledge.
Recent developments in the field of reinforcement learning show that RL techniques can replace or supplement model-based techniques~\cite{ernst2009reinforcement}.
A number of recent papers provide examples of how a popular RL method, Q-learning~\cite{sutton1998reinforcement}, can be used for demand response~\cite{oNeill2010residential,henze2003evaluation,kara2012using,PowellBias}.  
For example in~\cite{oNeill2010residential}, O'Neill \textit{et al.} propose an automated energy management system based on Q-learning that learns how to make optimal decisions for the consumers. 
In~\cite{henze2003evaluation}, Henze \textit{et al.} investigate the potential of Q-learning for the operation of commercial cold stores and in~\cite{kara2012using}, Kara \textit{et al.} use Q-learning to control a cluster of thermostatically controlled loads.
In~\cite{PowellBias}, Lee \textit{et al.} propose a bias-corrected form of Q-learning to operate battery charging in the presence of volatile prices.
While being a popular method, one of the fundamental drawbacks of Q-learning is its inefficient use of experiences, given that Q-learning discards the current data sample after every update.
As a result, more observations are needed to propagate already known information through the state space. 
In order to overcome this drawback, batch RL techniques~\cite{fonteneau2013batch,adam2012experience,ernst2005tree} can be used.
In batch RL a controller estimates a control policy based on a batch of experiences.
These experiences can be a fixed set~\cite{ernst2005tree} or can be gathered online by interacting with the environment~\cite{lange2012batch}.
Given that batch RL algorithms can reuse past experiences, they converge faster compared to techniques like Q-learning or SARSA.
This makes batch RL techniques suitable for practical implementations, such as demand response.
For example, the authors of~\cite{wen2014optimal} combine Q-learning with eligibility  traces in order to learn the consumer and time preferences of demand response applications.
In~\cite{RuelensBRLCluster}, the authors use a batch RL technique to schedule a cluster of electric water heaters and in~\cite{StijnBRL}, Vandael \textit{et al.} use a batch RL technique to find a day-ahead consumption plan of a cluster of electric vehicles.
An excellent overview of batch RL methods can be found in~\cite{lange2012batch} and~\cite{busoniu2010reinforcement}.

Inspired by the recent developments in batch RL, in particular fitted Q-iteration by Ernst \textsl{et al.}~\cite{ernst2009reinforcement}, this paper builds upon the existing batch RL literature and contributes to the application of batch RL techniques to residential demand response. 
The contributions of our paper can be summarized as follows:
	(1) we demonstrate how fitted Q-iteration can be extended to the situation when a forecast of the exogenous data is provided;
	(2) we propose a policy adjustment method that exploits general expert knowledge about monotonicity conditions of the control policy;
	(3) we  introduce a model-free Monte Carlo estimator method to find a day-ahead consumption plan by making use of a novel metric based on Q-values.

This paper is structured as follows: 
Section~\ref{building_blocks} defines the building blocks of our  batch RL controller.
Section~\ref{sec.mdp} formulates the problem as a Markov decision process. 
Section~\ref{sec.algorithms} describes our model-free batch RL techniques for demand response.
Section~\ref{experiments} demonstrates the presented techniques in a realistic demand response setting.
To conclude, Section~\ref{sec.results} summarizes the results and discusses further research.

\vspace{-2mm}
\section{Building blocks: model-free approach}
\label{building_blocks}
\begin{figure}[t!]
\centerline{\includegraphics[width=1\columnwidth]{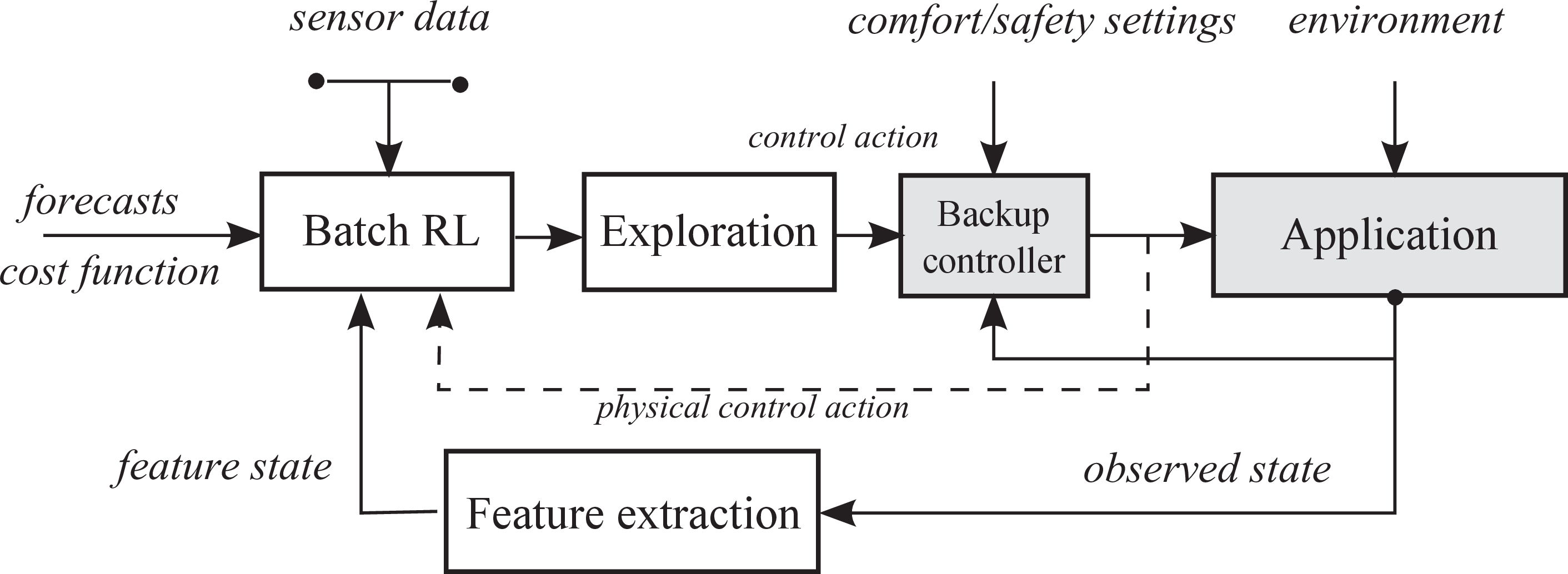}}
\caption{Building blocks of a batch Reinforcement Learning (RL) controller in a demand response application.}
\label{controlSchema}
\end{figure}
Fig.~\ref{controlSchema} presents a general overview of the different building blocks of our batch RL approach applied to a demand response setting.
This paper focuses on  two types of  thermostatically controlled loads.
The first type is a residential electric water heater with a stochastic hot-water demand~\cite{jordan2001realistic}.
The dynamic behavior of the electric water heater, used in this paper, is modeled by making use of a nonlinear stratified thermal tank model as described in~\cite{vanthournout2012smart}.
Our second demand response application is a heat-pump thermostat for a residential building.
The temperature dynamics of the building are modeled using a second-order equivalent thermal parameter model~\cite{chassin2008gridlab}, describing the temperature dynamics of the indoor air and of the building envelope. 
In order to develop a practical implementation we assume that the temperature of the building envelope is  a hidden state variable, and thus cannot be measured.
In addition, we assume that both  applications are equipped with a backup controller that guarantees the comfort and safety settings of the end users. 
The backup controller can be a built-in overrule mechanism that turns the application ON or OFF depending on the current state and a predefined switching logic.
The operation and settings of the backup controller are assumed to be unknown, however, the batch RL controller can measure the action of the backup controller~(see dashed arrow in Fig.~\ref{controlSchema}).

Before the observed state information of the demand response application can be sent to the batch RL algorithm, we apply a feature extraction technique~\cite{bertsekas1996neuro}.
A first task of the feature extraction technique is to extract non-observable state information that is required to obtain a policy.
For example, in our  implementation of a heat-pump thermostat we use  feature extraction  to represent the temperature of the building envelope, which cannot be measured.
A second task of the feature extraction technique could be to find a compact representation of the observable state.
For example, in the case of an electrical water boiler,   feature extraction is used to find a compact representation of the observable state.

At the start of each day the batch RL controller constructs a control policy for the next day, given a fixed batch of transitions and cost values.
The batch RL controller needs no a priori information on the model dynamics and considers its environment  a black box.  
As a result, the batch RL controller can be applied to virtually every demand response problem or even for cluster control~\cite{RuelensBRLCluster}.
During the day, an exploration strategy is used online to interact with the environment and to collect new transitions that are added systematically to the batch.

The goal of this paper is to develop a model-free controller for two relevant demand response business models~\cite{belpex,Dupont}: dynamic pricing and day-ahead scheduling.
The objective of the first business model is to adapt the consumption profile in response to an external price signal without violating the comfort settings of the end user.
The optimal solution is a closed-loop control policy that is a function of the current and past measurements of the state.
The second business model relates to the participation in the day-ahead market.
The objective is to construct the day-ahead consumption plan and then follow it during the day.
The goal is to minimize the cost in the day-ahead market and minimize any deviation between the day-ahead consumption plan and the actual consumption.
In contrast to the solution of the first business model, the day-ahead consumption plan is a feed-forward plan for the next day, i.e. an open-loop policy, which does not depend on measurements of the state.


\vspace{-1mm}
\section{Markov decision process formulation}
\label{sec.mdp}
In order to use reinforcement learning techniques, we formulate the sequential decision problem of a  demand response application as  a Markov decision process~\cite{BellmanDP,bertsekas1996neuro}.
The Markov decision process, used in this paper, is defined by its $d$-dimensional state space $X \subset \mathbb{R}^{d}$, its action space $U \subset \mathbb{R}$, its stochastic discrete-time transition function $f$, and its cost function $\rho$ \cite{fonteneau2013batch}.
The optimization horizon is considered finite, comprising $T \in \mathbb{N}\setminus \{0\}$ steps, where at each discrete time step $k$, the state evolves as follows:
\begin{equation}
x_{k+1} = f(x_{k},u_{k},w_{k}) ~~ \forall k \in \{1,\cdots,T-1\},
\label{eq.f}
\end{equation} with $w_{k}$  a realization of a random process drawn from a conditional probability  distribution $\probwk$,
 $u_{k} \in U$ the control action, and $x_{k} \in X$ the state.
Associated with each state transition, a cost $c_{k}$ is given by: 
\begin{equation}
c_{k}=\rho(x_{k},u_{k},w_{k}) ~~ \forall k \in \{1,\cdots,T\}.
\label{eq.reward}
\end{equation}
The goal is to find a  control policy ${h^{*}:X\rightarrow U}$ that minimizes the expected $T$-stage return for any state in the state space.
The expected $T$-stage return starting from $x_{1}$ and following $h^{*}$ is defined as follows:
\begin{equation}
J^{h^{*}}_{T}(x_{1}) = \underset{w_{k}\sim\probwk}{\expected} \left[  \sum_{k=1}^{T}{\rho(x_{k},h^{*}(x_{k}),w_{k})}\right].
\end{equation}
A convenient way to  characterize the  policy $h^{*}$ is  by using  a state-action value function or Q-function:
\begin{align}
Q^{h^{*}}(x,u) = \underset{w\sim\probw}{\expected} \left[\rho(x,u,w) +J^{h^{*}}_{T}(f(x,h^{*}(x),w)) \right].
\label{Qfunction}
\end{align}
The Q-function is the cumulative return starting from  state $x$, taking action $u$, and following $h^{*}$ thereafter.
Starting from a Q-function for every state-action pair, the policy is calculated as follows:
\begin{equation}
h^{*}(x)  \in \underset{u \in U}{\text{arg min~}} Q^{h^{*}}(x,u),
\label{Qpolicy}
\end{equation}
where $h^{*}$ satisfies the Bellman equation~\cite{BellmanDP}.
The next paragraphs give a formal description of the state space, the backup controller, and the cost function tailored to  demand response.

\subsection{State description}
The state space $X$ is spanned by a time-dependent state space component  $X_{\text{t}}$, a controllable state space component $X_{\text{ph}}$, and  an uncontrollable exogenous state space component $X_{\text{ex}}$~\cite{bertsekas1996neuro}:
\begin{equation}
X = \Xtime \times \Xphys \times \Xexo.   
\label{eq.statespace}
\end{equation}
\subsubsection{Timing}
The state space component $\Xtime$ describes the part of the state space related to timing, i.e. it carries timing information that is relevant for the dynamics of the system:
\begin{equation}
X_{\mathrm{t}} = \Xtimequarter \times \Xtimeday ~\mathrm{with}~ \Xtimequarter = \left\{1,...,96\right\}, \Xtimeday = \left\{1,...,7\right\},
\label{eq.timestate}
\end{equation}
where $x_{\mathrm{t}}^{\mathrm{q}} \in  \Xtimequarter $ denotes the quarter in the day, and $x_{\mathrm{t}}^{\mathrm{d}} \in  \Xtimequarter$ denotes the day in the week.
The rationale is that most consumer behavior tends to be repetitive and follows a diurnal pattern.  
\subsubsection{Physical representation}
The controllable state space component $X_{\mathrm{ph}}$ represents the physical state information related to the quantities that are measured locally and that are influenced by the control actions, e.g. the indoor air temperature or the state of charge of an electric water heater:
\begin{equation}
\xphys \in X_{\mathrm{ph}}~\mathrm{with}~ \xphysmin<\xphys<\xphysmax, 
\label{eq.physstate}
\end{equation} 
where $\xphysmin$ and $\xphysmax$ denote the lower and upper bound, set to guarantee the comfort and safety of the end user. 
\subsubsection{Exogenous Information}
The state description of the uncontrollable exogenous state is split into two components:
\begin{equation}
X_{\mathrm{ex}} = X_{\mathrm{ex}}^{\mathrm{ph}} \times X_{\mathrm{ex}}^{\mathrm{c}}.
\label{eq.state space}
\end{equation}
When  the random disturbance $w_{k+1}$ is independent of $w_{k}$, given $x_{k}$ there is no need to include an uncontrollable exogenous state information in the state space.
However, most physical processes, such as the outside temperature and solar radiation, exhibit a certain degree of autocorrelation, where the next state depends on the previous states.
For this reason we include an exogenous state space component  $ x_{\mathrm{ex}}^{\mathrm{ph}} \in X_{\mathrm{ex}}^{\mathrm{ph}}$ in our state space description~\cite{bertsekas1996neuro}.
This exogenous state space component is related to the observable exogenous information that has an impact on the physical dynamics and cannot be influenced by the  control actions.
The second exogenous state space component $  x_{\mathrm{ex}}^{\mathrm{c}} \in  X_{\mathrm{ex}}^{\mathrm{c}}$ has no direct influence on the dynamics, but contains information to calculate the cost $c_{k}$.
This work assumes that a deterministic forecast of the exogenous state information related to the cost $\lambda \in \mathbb{R}^{T}$ is provided for the time span covering the optimization problem.

\subsection{Backup controller}
\label{subsection.backup_controller}
In order to develop a practical demand response technology we assume that each device is equipped with an overrule mechanism that guarantees comfort and safety constraints. 
The backup function  ${B:X \times U \longrightarrow \Uphys}$ maps the requested control action $u_{k} \in U$ taken in  state $x_{k}$ to a physical control action $\uphk \in \Uphys$: 
\begin{equation}
\uphk= B(x_{k},u_{k}).
\end{equation} 
The settings of the backup function $B$ are unknown by the batch RL controller, but the resulting action  $\uphk$  can be measured (see dashed arrow in Fig.~\ref{controlSchema}).

\subsection{Cost function}
In general, RL techniques do not require detailed knowledge of the cost function.
However, for most demand response business models a cost function is available. 
This paper considers two typical cost functions related to demand response.
In the dynamic pricing scenario an external price profile $\lambda \in \mathbb{R}^{T}$ is known deterministically at the start of the horizon. 
The cost function is described as:
\begin{equation}
c_{k}=  \uphk\lambda_{k}\Delta t,
\label{eq_ToU}
\end{equation}
where $\lambda_{k}$ is the electricity price at time step $k$ and $\Delta t$ is the length of a control period.
The objective of the second business case is to determine a day-ahead consumption plan and to follow this plan during operation.
The day-ahead consumption plan should be minimized based on day-ahead prices. 
In addition, any deviation between the planned consumption and actual consumption should be avoided.
As such, the cost function can be written as:
\begin{equation}
c_{k}=  u_{k}\lambda_{k} \Delta t +  \alpha |u_{k}\Delta t -\uphk\Delta t |,
\label{day_ahead}
\end{equation}
where $u_{k}$ is the planned consumption, $\uphk$ is the actual consumption and $\lambda_{k}$ is the forecasted day-ahead price.
The first part of (\ref{day_ahead}) is the cost for buying energy at the day-ahead market.
The second part defines a penalty for any deviation between the planned consumption and the actual consumption.

\subsection{Reinforcement learning for demand response}
When the description of the transition function and cost function is available, techniques that make use of the  Markov decision process framework, such as approximate dynamic programming~\cite{powell2007approximate} or direct policy search~\cite{busoniu2010reinforcement}, can be used to find near-optimal policies.
However, in our implementation we assume that the transition function $f$, the backup controller $B$, and the underlying probability of the exogenous information are unknown. 
For this reason, we present a model-free batch RL approach that builds on previous theoretical work on RL, in particular fitted Q-iteration~\cite{ernst2005tree}, expert knowledge~\cite{busoniu2010exploiting}, and the synthesis of artificial trajectories~\cite{fonteneau2013batch}.

\vspace{-2mm}
\section{Algorithms}
\label{sec.algorithms}
Typically batch RL  techniques construct policies based on a batch of tuples of the form: $\batchCost$,
where  $ x_{l} = (x_{l,\mathrm{t}}^{\mathrm{q}} ,x_{l,\mathrm{t}}^{\mathrm{d}},\xlphys, x_{l,\mathrm{ex}}^{\mathrm{ph}})$ denotes the  state at time step $l$ and ${x}_{l}'$ denotes the state at time step $l+1$.
However, for most demand response applications, the cost function $\rho$ is given a piori, and of the form $\rho(x_{l},\uphl,\lambda)$.
As such, this paper considers tuples of the form $\tuple$.

\subsection{Fitted Q-iteration using a forecast of the exogenous data}
Here we  show how fitted Q-iteration~\cite{ernst2005tree} can be extended to the situation when a forecast  of the exogenous state space component  is provided (Algorithm~\ref{forecastedFQI}).
The algorithm iteratively builds a training set $\mathcal{T}_{\mathrm{reg}}$ with all  state-action pairs $(x,u)$ in $\mathcal{F}$  as the input.
The target values consist of the corresponding cost values $\rho(x,\uph,\lambda)$ and the optimal Q-values, based on the approximation of the  Q-function of the previous iteration, for the next states $\underset{u \in U}{\text{min~}}\widehat{Q}_{N-1}(\hat{x}_{l}',u) $.
For a finite horizon problem the stopping criterion is reached when $N=T$, where $T$ is the number of control periods in the optimization horizon and $N$ denotes the iteration.
It is important to note that $\hat{x}_{l}'$ denotes the successor state in $\mathcal{F}$, where the observed exogenous state space information $x_{l,\mathrm{ex}}^{\mathrm{ph~\prime}}$ is replaced by its forecasted value $\hat{x}_{l,\mathrm{ex}}^{\mathrm{ph}~\prime}$ (line~\ref{state_forecast} in Algorithm~\ref{forecastedFQI}). 
Note that  in our algorithm the next state contains information on the forecasted exogenous data, whereas for standard fitted Q-iteration~\cite{ernst2005tree} the next state contains past observations of the exogenous data.
By replacing the observed exogenous part of the next state by its forecasted value, the Q-function of the next state assumes that the exogenous information will follow its forecast.
\begin{algorithm}[t]
\caption{Fitted Q-iteration using a forecast of the exogenous data (extended FQI)}
\label{forecastedFQI}
\begin{algorithmic}[1] 
\algsetup{linenosize=\tiny}
\renewcommand{\algorithmicrequire}{\textbf{Input:}}
\REQUIRE $ \batchUphys, \{\hat{x}_{l,\mathrm{ex}}^{\mathrm{ph}}\}_{l=1}^{\#\mathcal{F}},\lambda $\\
\STATE $N\leftarrow 0$
\STATE let $\widehat{Q}_{0}$ be zero everywhere on $X$ $\times$ $U$
\REPEAT
\FOR {$l = 1,\cdots,\#\mathcal{F}$}
\STATE $~~c_{l} \leftarrow \rho(x_{l},\uphl,\lambda)$  
\STATE $\text{~~}\hat{x}_{l}'\leftarrow (x_{l,\mathrm{t}}^{\mathrm{q}~\prime},x_{l,\mathrm{t}}^{\mathrm{d}~\prime},\xlphys^{~\prime},\hat{x}_{l,\mathrm{ex}}^{\mathrm{ph}~\prime})$ \label{state_forecast} $\vartriangleright$ \textit{replace~the \\ \text{~~}observed exogenous part of the next state~$x_{l,\mathrm{ex}}^{\mathrm{ph~\prime}}$\\ \text{~} by its forecasted value $\hat{x}_{l,\mathrm{ex}}^{\mathrm{ph}~\prime}$}
\STATE $~~Q_{N,l}\leftarrow c_{l} +\underset{u \in U}{\text{min~}}\widehat{Q}_{N-1}(\hat{x}_{l}',u)  $  
\ENDFOR
\STATE use regression to obtain $\widehat{Q}_{N}$ from \text{~~~~~~~~~~~~~~~~~~~~} $\mathcal{T}_{\mathrm{reg}}=\left\{\left((x_{l},u_{l}),Q_{N,l}\right),l =1,\cdots,\#\mathcal{F}\right\}$
\STATE increment $N$
\UNTIL{stopping criterion is reached}
\ENSURE $Q^{*}=\widehat{Q}_{N}$
\end{algorithmic}
\end{algorithm}
The proposed algorithm is relevant for demand response applications that are influenced by exogenous weather data.
Examples of these applications are heat-pump thermostats and air conditioning units.

In principle, any regression algorithm, such as neural networks~\cite{riedmiller2005neural}, can be applied in combination with fitted Q-iteration.
However, because of their robustness and fast calculation time, an extremely randomized trees ensemble method~\cite{ernst2005tree} is used.

\subsection{Expert policy adjustment}
Given the Q-function from Algorithm~\ref{forecastedFQI}, a near-optimal policy can be constructed by solving~(\ref{Qpolicy}) for every state in the state space.
In this section, we show how  expert knowledge on the monotonicity  of the policy can  be exploited to regularize the policy.
The  method enforces  monotonicity conditions by using a  convex optimization to approximate the policy, where expert knowledge is included in the form of extra constraints.
These constraints can result directly from the expert or  from a model-based solution.
In order to define a convex optimization problem we use a fuzzy model with triangular membership functions~\cite{busoniu2010reinforcement} to approximate the policy.
The centers of the triangular membership functions are located on an equidistant grid with $N_{\mathrm{g}}$ membership functions  along each dimension of the state space. 
This partitioning leads to $N_{\mathrm{g}}^d$ state-dependent membership functions for each action. 
The parameter vector $\theta^{*}$ that approximates the original policy can be found by solving the following least-squares problem:
\begin{equation}
\begin{aligned}
\theta^{*} \in\text{~}&\underset{{\theta}}{\text{~arg min}}\sum_{l=1}^{\#\mathcal{F}}{\Big([F(\theta)](x_{l})-h^{*}(x_{l})\Big)^{2}}, \\
&\text{s.t. expert knowledge}
\label{leastSquareProblem}
\end{aligned}
\end{equation}
where $F$ denotes an approximation mapping of a  weighted linear combination of triangular membership functions and $[F(\theta)](x)$ denotes the policy $F(\theta)$ evaluated at state $x$.
Let $h^{*}$ be the policy obtained by solving~(\ref{Qpolicy}), given the the Q-function obtained by Algorithm~\ref{forecastedFQI}.
A more detailed description of how these triangular membership functions are defined can be found in~\cite{busoniu2010reinforcement}.
The fuzzy approximation of the policy allows us to add expert knowledge to the policy in the form of convex constraints of the least-squares problem defined in~(\ref{leastSquareProblem}), which can be solved using a convex optimization solver.
Using the same notation as in~\cite{busoniu2010exploiting},  we can enforce monotonicity conditions along the $d$th dimension of state space as follows:
\begin{equation}
\delta_{d} [F(\theta)](x_{d})  \leq  \delta_{d} [F(\theta)]({x'}_{d}) 
\end{equation}
for all state components $x_{d}\leq{x'}_{d}$ along the  dimension $d$. 
If $\delta_{d}$ is -1 then $[F(\theta)]$ will be decreasing along the $d$th dimension of $X$, whereas if $\delta_{d}$ is 1 then $[F(\theta)]$ will be increasing along the $d$th dimension of $X$.
Once $\theta^{*}$ is found, the adjusted policy $\hat{h}$, given this expert knowledge, can be calculated as ${\hat{h}(x) = [F(\theta^{*})](x)}$.
When the batch $\mathcal{F}$ contains a limited number of tuples, e.g. only a few days,  the expert policy adjustment method can be used to improve the quality of the policy of a demand response problem.

\subsection{Day-ahead consumption plan}
\label{sec.planning}
This section explains how to construct a day-ahead schedule starting from the Q-function obtained by Algorithm 1.
Finding  a day-ahead schedule  has a direct relation to two situations: 
1) a day-ahead market, where participants have to submit a day-ahead schedule one day in advance of the actual consumption~\cite{belpex};
2) a distributed optimization process, where two or more participants are coupled by a common constraint, e.g. congestion management~\cite{TRIANA}.
Algorithm~\ref{algoAT} describes a model-free Monte Carlo estimator method~\cite{fonteneau2013batch} for policy evaluation that makes use of a metric based on Q-values.
The  method estimates the average return of a  policy by synthesizing $p$ sequences of transitions of length $T$ from $\mathcal{F}$.
These $p$ sequences can be seen as a proxy of the actual trajectories that could be obtained by simulating the policy on the given control problem.
Note that since we consider a stochastic setting, $p$ needs to be greater than 1.
A sequence is grown in length by selecting a new transition among the samples of not-yet-used one-step transitions in $\mathcal{F}$.
Each new transition is selected by  minimizing a distance metric  with the previously selected transition.\\
\indent In~\cite{fonteneau2013batch}, Fonteneau \textit{et al}. propose the following distance metric in $X{\times}U: \Delta \left(\left(x,x'\right),\left(u,u'\right)\right) = \Vert x-x' \Vert+\Vert u-u' \Vert$, where $\Vert \cdot \Vert$ denotes the Euclidean norm.
It is important to note that  this metric weighs each  dimensions of the state space equally.
In order to overcome specifying  weights to each dimension, we propose a distance metric as specified on line~\ref{metriek} of Algorithm~\ref{algoAT}.
Here ${Q}^{*}$ is obtained by applying Algorithm~\ref{forecastedFQI} and $x_{k}^{i}$ denotes the state corresponding to the $i$th trajectory at time step $k$.
The artificial trajectory $P^{i}$ contains the control actions corresponding with the optimal Q-value, given the state $x_{k}^{i}$ (see line~\ref{artificial_trajectory}).
The next state $x_{l^{i}}'$ is found by taking the next state of the tuple that minimizes the distance metric (see line~\ref{metriek}).
The regularization parameter $\xi$  is a scalar that is included to penalize states that have similar Q-values, but have a large Euclidean norm in the state space. 
When the Q-function is strictly increasing or decreasing $\xi$  can be set to 0.
The motivation behind using Q-values instead of the Euclidean distance in $X \times U$ is that Q-values capture the dynamics of the system and, therefore, there is no need to select individual weights.
\begin{algorithm}[t!]
\caption{Model-free Monte Carlo method~\cite{fonteneau2013batch}.}
\label{algoAT}
\begin{algorithmic}[1] 
\algsetup{linenosize=\tiny}
\renewcommand{\algorithmicrequire}{\textbf{Input:}}
\REQUIRE $\batchUphys $, $\{\hat{x}_{l,\mathrm{ex}}^{\mathrm{ph}}\}_{l=1}^{\#\mathcal{F}} $,$\lambda$, ${x}_{1}$, $p$, $\xi$\\
\STATE $\mathcal{G}\leftarrow \mathcal{F}$
\STATE Apply Algorithm~\ref{forecastedFQI}, to obtain ${Q}^{*}$
\FOR {$i=1,\cdots,p$}
\STATE $k \leftarrow 1$
\STATE $x_{k}^{i} \leftarrow x_{1}$
  \WHILE {$k<T$}
  \STATE $u_{k}^{i} \leftarrow \underset{u'\in U}{\text{arg min~}} {Q}^{*}(x_{k}^{i},u')$ \label{artificial_trajectory} \\
	\STATE $\mathcal{H} \leftarrow \underset{(x_{l},u_{l},x_{l}',u_{l}^{\mathrm{ph}}) \in \mathcal{G} }{\text{arg min}}  \vert {Q}^{*}(x^{i}_{k},u^{i}_{k})-{Q}^{*}(x_{l},u^{i}_{k}) \vert + \text{~~~~~~~~~~~~~~~~~~~~~~~~~~~~~~~~~~~~~~~~~~~~}\xi \Vert x_{k}^{i} - x_{l} \Vert$ \label{metriek}
  \STATE $l^{i} \leftarrow$ lowest index in $\mathcal{G}$ of the transitions in  $\mathcal{H}$
	\STATE $P_{k}^{i} \leftarrow u_{k}^{i}$
			\STATE $k \leftarrow k+1$
  \STATE $x_{k}^{i} \leftarrow x_{l^{i}}'$
	\STATE $\mathcal{G}\leftarrow \mathcal{G} \backslash \left\{(x_{l^{i}},u_{l^{i}},x_{l^{i}}',u_{l^{i}}^{\mathrm{ph}})\right\}$
  \ENDWHILE
\ENDFOR
\ENSURE $P^{1},\cdots,P^{p}$
\end{algorithmic}
\end{algorithm}

\vspace{-2mm}
\section{Simulations}
\label{experiments}
This section presents the simulation results of three experiments and evaluates the performance of the proposed algorithms.
We focus on two examples of flexible loads, i.e. an electric water heater and heat-pump thermostat.
The first experiment  evaluates  the performance of extended FQI (Algorithm 1)  for a heat-pump thermostat.
The rationale behind using extended FQI for a heat-pump thermostat, is that the temperature dynamics of a building is influenced by exogenous weather data, which is not the case for an electric water heater.
In the second experiment, we apply the policy adjustment method to an electric water heater. 
The final experiment uses the model-free Monte Carlo method  to find a day-ahead consumption plan for a heat-pump thermostat.
It should be noted that the policy adjustment method and model-free Monte Carlo method can also be applied to both heat-pump thermostat and electric water heater.

\subsection{Thermostatically controlled loads}
\label{TCLs}
Here we describe the state definition and the settings of the backup controller of the electric water heater and the heat-pump thermostat.

\subsubsection{Electric water heater}
We consider that the storage tank of the electric water heater is equipped with $n_{\mathrm{s}}$ temperature sensors.
The full state description of the electric water heater is  defined as follows:
\begin{equation}
x_{k} = (x_{k,t}^{\mathrm{q}},T^{1}_{k},\cdots,T^{i}_{k},\cdots,T^{n_{\mathrm{s}}}_{k}),\\
\label{eq.DHN}
\end{equation}
where $\xkt$ denotes of the current quarter in the day  and $T^{i}_{k}$ denotes the temperature measurement of the $i$th sensor.
This work uses a feature extraction to reduce the dimensionality of the controllable state space component by replacing it with
with the average sensor measurement. 
As such the reduced state is defined as follows:
\begin{equation}
x_{k}  = (\xkt,\frac{ \sum_{i=1}^{n_{\mathrm{s}}} {T}^{i}_{k}}{n_{\mathrm{s}}}).
\label{ewhstate}
\end{equation}
More generic dimension reduction techniques, such as an auto-encoder network and a principle components analysis~\cite{hinton2006reducing,lange2010deep} will be explored in future research.
The logic of the backup controller of the electric water heater is defined as: 
\begin{align}
B(x_{k},u_{k}) = \left\{\begin{matrix}
\uphk{=}&u_{\mathrm{max}}&\text{if }& {\soc \leq}30\%\\ 
\uphk{=}&u_{k}&\text{if }& {30\%}{<\soc<}100\%.\\ 
\uphk{=}&0 &\text{if }& {\soc\geq} 100\%
\end{matrix}\right.
\end{align} 
The electric heating element of the electric water heater can be controlled with a binary control action $u_{k} \in \{0,\umax\}$, where $\umax=2.3$kW is the maximum power.
A detailed description of the nonlinear dynamics and calculation of the state of charge $\soc$ is out of the scope of this paper and can be found in~\cite{vanthournout2012smart}.
The stratified thermal tank model consists of 50 layers of which 8 are measured to construct the full state~\cite{vanthournout2012smart}.
We use a set of hot-water demand profiles with a mean load of 100~l/day obtained from~\cite{jordan2001realistic} to simulate a realistic tap demand.

\subsubsection{Heat-pump thermostat}
Our second application considers a heat-pump thermostat that  can measure the indoor air temperature, the outdoor air temperature, and  the solar radiation.
The full state of the heat-pump thermostat is defined as follows:
\begin{equation}
x_{k}=(\xkt,\Tkin,\Tkout,S_{k}),
\end{equation}
where the physical state space component consists of the indoor air temperature $\Tkin$ and the exogenous state space component  contains the outside air temperature $\Tkout$ and the solar radiation $S_{k}$.
The internal heat gains, caused by user behavior and electric appliances, cannot be measured and are not included in the state.
We included a measurement of the solar radiance in the state since
solar energy transmitted through windows can significantly impact the indoor temperature dynamics.
In order to have a practical implementation, we consider that we cannot measure the temperature of the building envelope.
Similar to~\cite{bertsekas1996neuro}, we used a feature extraction technique based on past state observations by including a virtual building envelope temperature, which is  a running average of the past $n_{\mathrm{r}}$ air temperatures:
\begin{equation}
x_{k}     = (\xkt,\Tkin,\frac{\sum_{i=k-n_{\mathrm{r}}}^{k-1} T_{i,\mathrm{in}}}{n_{\mathrm{r}}},\Tkout,S_{k}).
\label{eq_StateAirco}
\end{equation}
An alternative method could be to use a model-based technique to estimate the temperature of the building envelope.
The backup controller of the heat-pump thermostat is defined as follows:
\begin{equation}
B(x_{k},u_{k}) = \left\{\begin{matrix}
\uphk=& u_{\mathrm{max}} &\text{if~}& \Tkin \leq \underline{T}_{k} \\ 
\uphk=& u_{k}    &\text{if~}&\underline{T}_{k} < \Tkin <\overline{T}_{k},\\ 
\uphk=& 0          &\text{if~}&  \Tkin \geq \overline{T}_{k} 
\end{matrix}\right.
\end{equation} 
where $\underline{T}_{k}$ and $\overline{T}_{k}$ are the minimum and maximum temperature  settings defined by the end user.
The action space of the heat pump is discretized in 10 steps, $u_{k} \in \{0,...,\umax\}$, where $\umax=3$kW is the maximum power.
In the simulation section we define a minimum and maximum comfort setting of $19^{\circ}\mathrm{C}$ and $23^{\circ}\mathrm{C}$.
A detailed description of the temperature dynamics of the indoor air and building envelope can be found in~\cite{chassin2008gridlab}.
The exogenous information consists of the outside temperature, the solar radiation,  and the internal heat gains from a location in Belgium~\cite{crawley2001energyplus,Dupont}.

In the following experiments we define a control period of one quarter and an optimization horizon $T$ of 96 control periods.

\subsection{Experiment 1}
The goal of the first experiment is to compare the performance of  Fitted Q-Iteration (standard FQI)~\cite{ernst2005tree} to the performance of our extension of FQI (extended FQI, given by Algorithm~\ref{forecastedFQI}).
The objective of the considered heating system is to minimize the electricity cost of the heat pump by responding to an external price signal. 
The electricity prices are taken from the Belgian wholesale market~\cite{belpex}. 
We assume that a forecast of the outside temperature and solar radiance is available.
Since the goal of the experiment is to assess the impact on the performance of  FQI when a forecast is included, we assume perfect forecasts of the outside temperature and solar radiance.
The observed state information of the controller is defined by (\ref{eq_StateAirco}) with $n_{\mathrm{r}}$ set to 3, and the cost function is given by (\ref{eq_ToU}).
\begin{figure}[t!]
\centerline{\includegraphics[width=0.9\columnwidth]{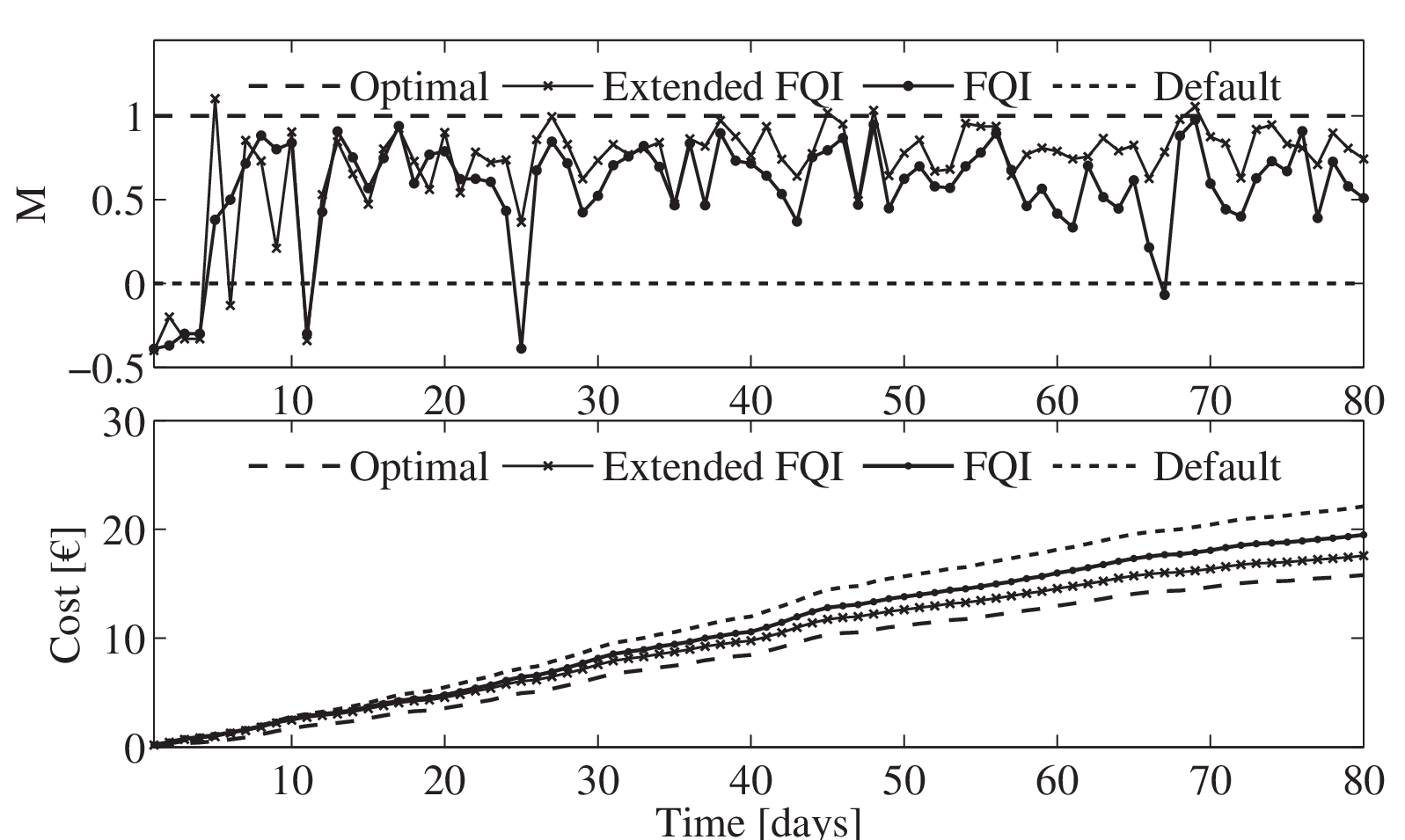}}
\caption{Simulation results for a heat-pump thermostat and a dynamic pricing scheme using an optimal controller, Fitted Q-Iteration (FQI) with and without forecast,  and a default controller.
The top plot depicts the performance metric $M$ and the bottom plot depicts the cumulative electricity cost.}
\label{fig_projected}
\end{figure}
During the day, both FQI controllers use an $\varepsilon$-greedy exploration strategy. 
This exploration strategy selects a random control action with probability ${\varepsilon}_k$ and follows the policy with probability $1-{\varepsilon}_k$.
The exploration probability ${\varepsilon}_k$ is decreased on a daily basis following a harmonic sequence~\cite{powell2007approximate}. 
At the end of each day the FQI controller adds the tuples of the previous day to the batch and computes a policy for the next $T$ time steps~\cite{gabel2011improved}.\\
\indent In order to compare the performance of the FQI controllers we define the following metric:
\begin{equation}
M = \frac{c_{\mathrm{fqi}} - c_{\mathrm{dc}}}{c_{\mathrm{oc}} - {c}_{\mathrm{dc}}},
\end{equation}
where $c_{\mathrm{fqi}}$ denotes the daily cost of the FQI controller, $c_{\mathrm{dc}}$ denotes the daily cost of the default controller and  $c_{\mathrm{oc}}$ denotes the daily cost of the optimal controller.
The metric $M$ corresponds to 0 if the FQI controller obtains the same performance as the default controller and corresponds to 1 if the FQI controller obtains the same performance as the optimal controller.
The default controller is a  hysteresis controller that switches on when the indoor air temperature is lower than 19${}^{\circ}\mathrm{C}$ and stops heating when the indoor air temperature reaches 20${}^{\circ}\mathrm{C}$.
The optimal controller is a model-based controller that has full information on the model parameters and has perfect forecasts of all exogenous information.\\
\indent The simulation results of the heating system  for a simulation horizon of 80 days are depicted in Fig.~\ref{fig_projected}.
The top plot depicts the daily metric $M$ and the bottom plot depicts the cumulative electricity cost of the heat-pump thermostat.
The average metric $M$ over the simulation horizon is $0.56$ for standard FQI and $0.71$ for extended FQI, which is an improvement of $27\%$.
The performance gap of $0.29$ between  extended FQI and the optimal controller is a reasonable result given that the model dynamics and disturbances are unknown, and that  exploration is included.\\
\indent Extended FQI  was able to decreases the total electricity cost with $19\%$ compared to the default controller over the total simulation horizon, whereas standard FQI decreased the total electricity cost with $14\%$.
It is important to note that the reduction of $19\%$ is not a result of lower energy consumption, since the energy consumption increased with $4\%$ compared to the default controller when extended FQI was used. \\
\indent With these experiment we showed that we successfully extended fitted Q-iteration to incorporate a forecasts of the exogenous data.

\subsection{Experiment 2}
\begin{figure}[t!]
\centerline{\includegraphics[width=1\columnwidth]{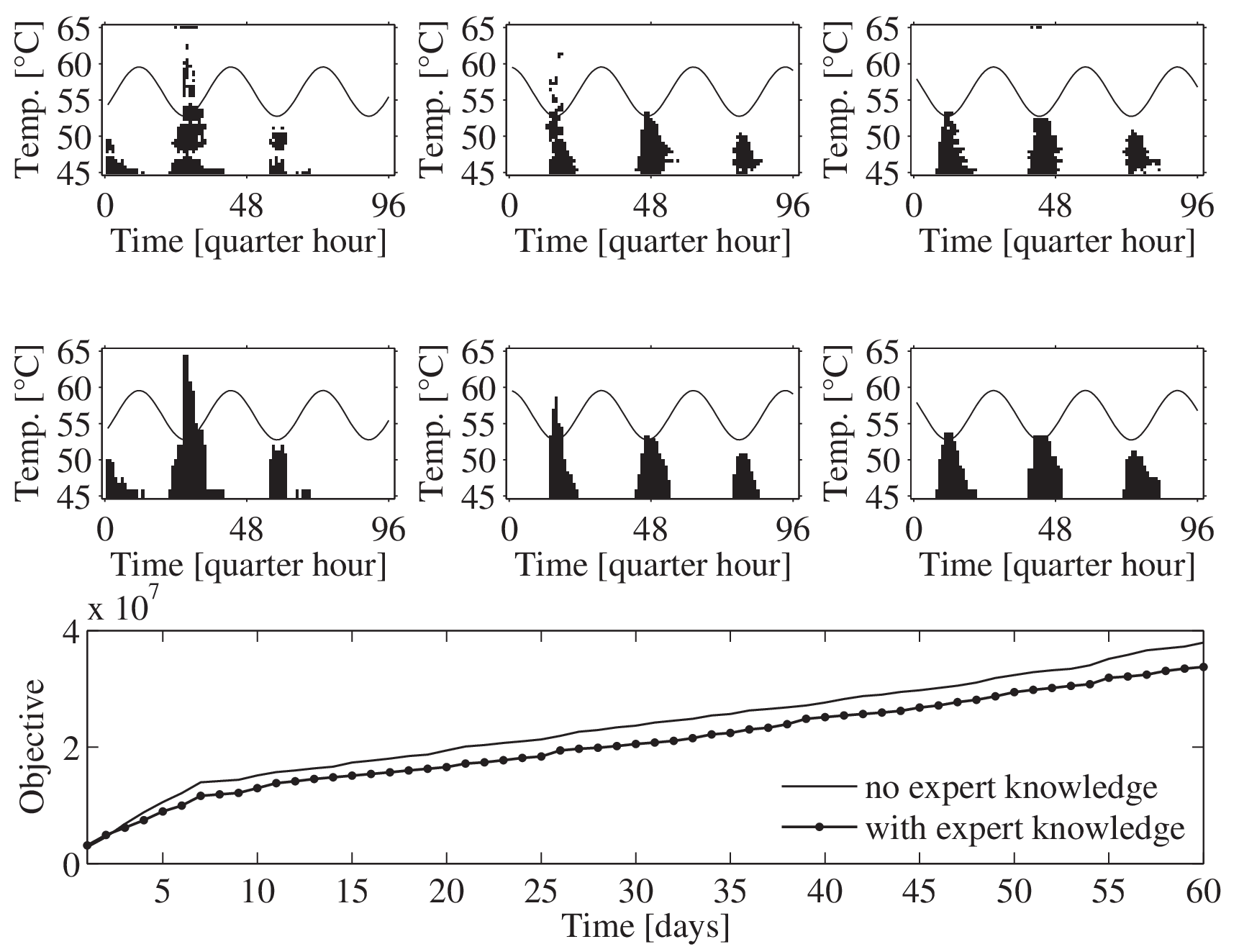}}
\caption{Simulation results for an electric water heater and dynamic pricing scheme. 
The top depicts the original  policies and  the middle row depicts the repaired policies for day 7, 14, and 21.  The price profile corresponding to each policy is depicted in the background.
The bottom plot illustrates the impact of adding expert knowledge on the objective.}
\label{fig_PolicyRegularization}
\end{figure}

The following experiment demonstrates the policy adjustment method for an electric water heater. 
As an illustrative example we used a sinusoidal price profile.
As stated in~(\ref{ewhstate}), the state space of an electric water heater consists of two dimensions, i.e. the time component and the average temperature.
Here we enforce a monotonicity constraint along the second dimension, which contains the average temperature.
The original policies after 7, 14 and 21 days obtained with  fitted Q-iteration can be seen in the top row of plots of Fig.~\ref{fig_PolicyRegularization}.
It can be seen that the original policies, obtained by fitted Q-iteration, violate the monotonicity constraints along the second dimension in several states.
The adjusted policies obtained by the policy adjustment method are depicted on the middle row of  plots. 
The simulation results indicate that when the number of tuples in the batch increases, the original and adjusted policies converge.
The bottom plot depicts the cumulative cost of fitted Q-iteration with and without expert knowledge over the simulation horizon.
The policy adjustment method was able to reduce the total objective by $11\%$ over 60 days.
The results conclude that when the number of tuples in $\mathcal{F}$ is small, the expert policy adjustment method can be used to improve the performance of standard fitted Q-iteration.

\subsection{Experiment 3}
\begin{figure}[t!]
\centerline{\includegraphics[width=0.95\columnwidth]{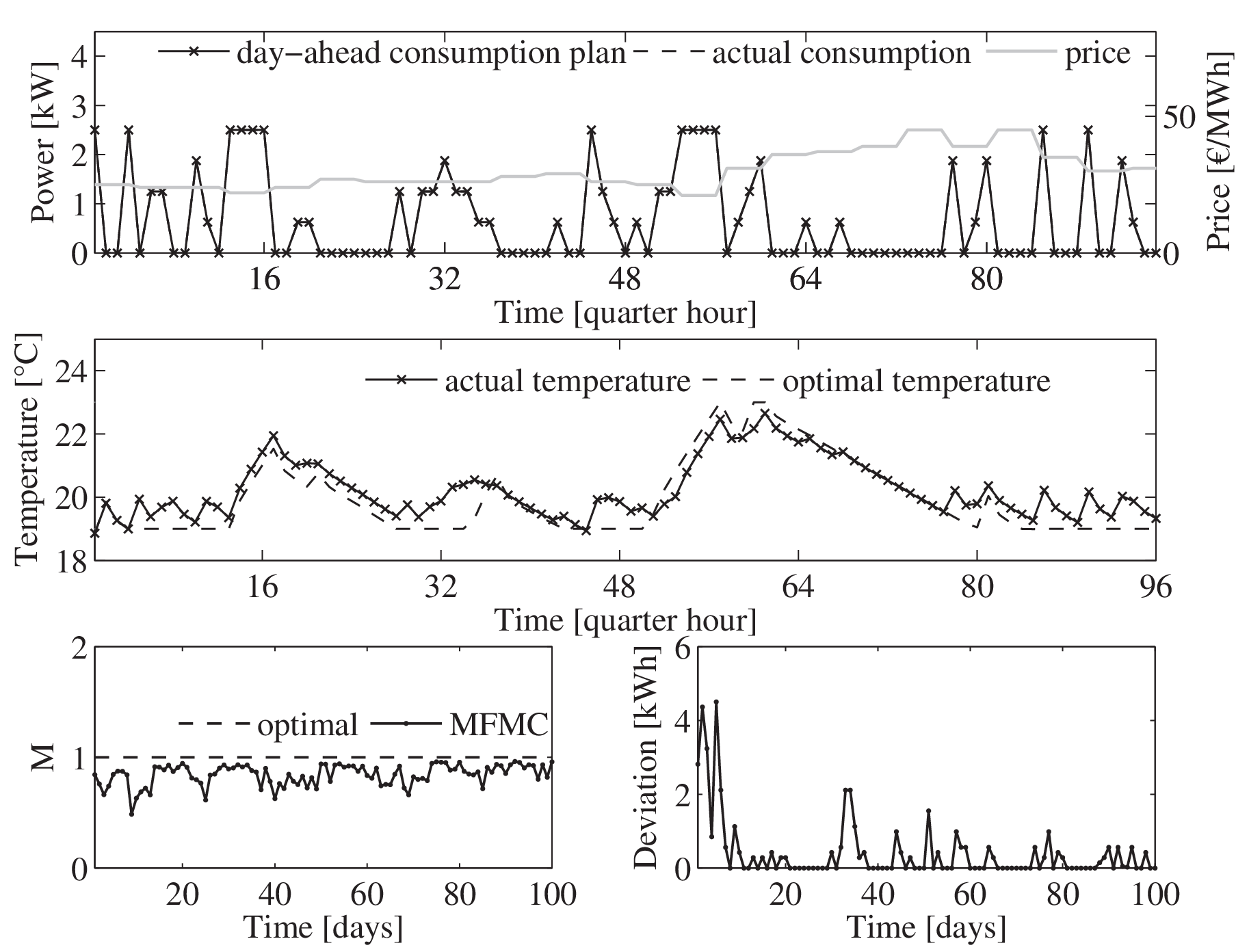}}
\caption{The top plot depicts the day-ahead consumption plan, actual consumption, and day-ahead price. The middle plot depicts the actual and optimal indoor temperature. The left bottom plot depicts the metric $M$. The daily deviation between the day-ahead consumption plan and actual consumption is given in the right bottom plot.}
\label{fig.heatPumpAT}
\end{figure}

The final experiment demonstrates the Model-Free Monte Carlo (MFMC) method (Algorithm~\ref{algoAT}) for finding the day-ahead consumption plan of a heat-pump thermostat.
The state variable is defined by (\ref{eq_StateAirco}), and the cost function is defined by (\ref{day_ahead}).
The parameter $\alpha$ was set to $10^3$ to penalize possible deviations between the planned consumption profile and the actual consumption.
The results of the experiment are depicted in Fig.~\ref{fig.heatPumpAT}. 
The parameter $p$, which indicates the number of artificial trajectories, was set to 4.
A day-ahead consumption plan was obtained by taking the average of these $4$ trajectories.
In order to assess the performance of the MFMC method, we defined an optimal controller that can exploit the model and has prescient knowledge on the internal heat gains.
The top plot depicts the day-ahead consumption plan and actual consumption profile of a mature controller that contains a batch of 60 days. 
The corresponding indoor air temperature of the presented controller and optimal controller is depicted in the middle plot.
In order to compare the daily cost of the MFMC method to the optimal controller, we define $M=c_{\mathrm{MFMC}}/c_{\mathrm{oc}}$, where $c_{\mathrm{MFMC}}$ is the daily cost of the MFMC method and $c_{\mathrm{oc}}$ is the daily cost of the optimal controller.
The metric $M$ for each day and  the deviation between the day-ahead consumption plan and the actual followed consumption are depicted in the bottom plot.
As the bottom plot indicates, the daily deviations decrease as the number of days in the batch increases.
The mean metric $M$ over the whole simulation period, including the exploration phase is $0.81$. 
The results of the last experiment imply that the model-free Monte Carlo method can be successfully used to construct a  forward consumption plan for the next day.

\vspace{-2mm}
\section{Conclusion}
\label{sec.results}
Driven by the challenges presented by the system identification step of model-based controllers, this paper contributed to the application of model-free batch Reinforcement Learning (RL) techniques to a demand response setting.
Motivated by the fact that some demand response applications, e.g  a heat-pump thermostat, are influenced by exogenous weather data, we adapted a standard batch RL technique, fitted Q-iteration, to incorporate a forecast of the exogenous data.
Numerical results have been presented that indicate that the proposed extension of fitted Q-iteration was able to improve the performance of standard fitted Q-iteration by $27\%$.
In general, batch RL techniques do not require any prior knowledge on the system behavior or the solution.
However, for some demand response applications, expert knowledge about the monotonicity of the solution, i.e. the policy, can be available.
As such, we presented an expert policy adjustment method that can exploit this expert knowledge. 
The results of an experiment with an electric water heater indicate that the policy adjustment method was able to reduce the cost objective by $11\%$ compared to fitted Q-iteration without expert knowledge.
A final challenge for model-free batch RL techniques is that of finding a day-ahead consumption plan, i.e. an open-loop solution.
In order to solve this problem we presented a model-free Monte Carlo method and we successfully tested the method for finding the day-ahead consumption plan of a heat pump.

Our future research in this area will focus on employing the presented algorithms in a realistic lab environment.
\vspace{-2mm}
\section*{Acknowledgments}
The authors would like to thank the reinforcement learning team from the University of Li\`ege for their  comments and suggestions.

\ifCLASSOPTIONcaptionsoff
  \newpage
\fi
\vspace{-6mm} 
\bibliographystyle{IEEEtran}  
 \small
\bibliography{references} 
\end{document}